\documentclass{elsart1p}
\usepackage{graphicx}
\usepackage{amssymb}
\begin{document}
\begin{frontmatter}
\title{QCD with and in nuclei: color transparency and short-range correlations in nuclei - 
theory, observations, directions for further studies}

\author{Mark Strikman}
\address{104 Davey Lab, Penn State University, University Park, PA 16802, USA}
\begin{abstract}
We summarize basic theoretical ideas which let to the observation of the short-range correlations (SRC) in  nuclei using hard probes  and  outline directions for probing quark-gluon structure of SRCs. 
Implications of the observations of color transparency for processes involving  pions are reviewed. Open questions  and  directions for further studies of color transparency phenomena  using hadronic projectiles are presented using as an  example the PANDA detector at FAIR.

\end{abstract}
\end{frontmatter}
%
\section{Short-range correlations in nuclei}
\label{src}
The presence and structure of high momentum/ short-distance short-range correlations (SRC) in nuclei were 
actively discussed in the 60's. However by  70's a consensus
has emerged  that it is impossible to observe them experimentally since effects of SRCs cannot be separated  from multistep processes. However we argued starting in 1975 that the use of the high energy processes where the energy and momentum transfer to the nucleon in the nucleus are much larger than the scale of momenta and energies in the SRCs will allow to resolve them. Two key features of SRC which can be tested directly using high energy probes are a) universality of the wave (spectral) functions 
of SRC which is due to dominance of the local singular interactions between two (three )... nucleons - 2N (3N,...) SRCs, b) emission of fast nucleons in the SRC decay following a fast removal of a nucleon from the SRC.

A power law decrease of the nucleon - nucleon potential at large nucleon momentum allowed to introduce a  decomposition of high momentum component of the nuclear wave function over contributions of 2N, 3N,... SRC  and to relate the corresponding contributions  to the high momentum distribution in the deuteron
\cite{FS77,FS81}.  The few nucleon SRC approximation 
 allowed to explain regularities of the processes of fast backward nucleon production - dominance of 2N SRC, importance at high momenta of the higher SRCs.
It was also suggested \cite{FS81} that study of inclusive (e,e') reactions at sufficiently large $Q^2 $ and $x > 1$ will allow to check universality of the SRCs via scaling of the ratios: $\sigma_A(x,Q^2)/\sigma_{^2H}(x,Q^2)$ for $1< x < 2$ and $\sigma_A(x,Q^2)/\sigma_{^3He}(x,Q^2)$ for $2< x < 3$. The analysis of the available data in \cite{Frankfurt:1993sp} confirmed presence of the scaling 
of the ratios for $1 < x < 2$ and predictions of \cite{FS81,FS88} for the values of the ratios and provided arguments for the cancelation  the f.s.i. for the ratios. Recently  a significant progress was made in the study of these ratios: the theory of the f.s.i. in the $x> 1$ kinematics was elaborated \cite{Frankfurt:2008zv} while 
 dedicated experiments at TJNAF confirmed presence of the scaling of ratios and extended it to the $x> 2$ region
\cite{Kim2,Kim1}, for a detailed review and references see \cite{Frankfurt:2008zv}.

We also  argued  that in order  to get a detailed information about the structure of SRC one needs to focus on semiexclusive reactions with large momentum transfer to the nucleon, like nucleon knockout in the $(e,e'p), (p, 2p)$ reactions.  Selection of the kinematics where the projectile interacts with a fast nucleon in the nucleus, and knocks it out with a momentum much larger than its initial momentum allows to study the decay function of the residual system with a clean separation between the knocked out nucleon and products of the decay of the SRC \cite{FS88}. The first experiment to look for such decays was performed at BNL \cite{eip2}. In the studied   kinematics  
incoming proton knocked out a  proton which initially had a forward 
momentum above $k_F$ and a backward neutron was detected. 
The analysis  \cite{eip3} has demonstrated that the process is dominated by the scattering off $pn$ SRCs in nuclei and that most of the protons are correlated with neutrons with $pp$ SRCs been a small correction.  This was confirmed by direct measurement of the ratio of $(e,e'pp)$ and $(e,e'pn) $ cross sections in the kinematics where a fast backward nucleon was knocked out and proton (neutron) from the decay was produced forward \cite{eip4,eip5}.

The momentum transfer, $-t \sim 5 \mbox{GeV}^2$ in the (p,2p)
process  was much larger than in the (e,e'pN) experiment 
($Q^2=2 {GeV}^2$). Also
the invariant mass of the produced $NN $ system was  much larger in the (p,2p) case, which strongly reduced the effects of the final state interaction between the nucleons. Consistency of two experiments provides an additional argument for the observation of the SRCs in these processes.
The dominance of $pn$ SRCs was discussed in the literature for a long time. Quantitative analyses of the double momentum distributions of   nuclei which reflect this feature of the decay function were presented in \cite{eheppn1,Schiavilla,Alvioli07}.

A joint analysis of the knock out and (e,e') reactions allows to conclude that (a) more than $\sim 90\%$ of  all nucleons with momenta $k\ge  \mbox{300 MeV/c}$  belong to two nucleon SRC correlations, (b) Probability for a given proton  with momenta $600\ge  k \ge \mbox{300 MeV/c}$ to belong to a $pn$ correlation is  $\sim$ 18 times larger than to belong to a $ pp$ correlation, (c)  Probability for a nucleon to have momentum $\ge$ 300 MeV/c in medium nuclei is  $\sim 25\% $, (d)  2N SRC mostly build of two nucleons with nonnucleonic configurations  (six quarks, $ \Delta N, \Delta \Delta$) constituting less than   $10 \div  20\%$  of the 2N SRCs, (e)
 Three nucleon SRC are present in nuclei with a significant probability.

Future (e,e')  experiments at $x > 1$ will   perform detailed 
studies of the  scaling of the ratios  and reach $Q^2$ where inelastic processes will lead to increase of the ratios. Semiexclusive reactions will allow to measure fine detailed of  2N SRCs, explore 3N SRCs, and look for the nonnucleonic degrees of freedom in SRCs (for example in the process $pA\to p\Delta^{++} (A-1)* $), for the recent summary see  \cite{Frankfurt:2008zv}. It would be very important to continue a parallel program of studies of these processes using electron and hadron beams. Though the interaction of the hadronic projectiles is somewhat more complicated, the large angle processes with proton (antiproton) projectiles allow an easier  selection of scattering off forward nucleons. They also  allow to reach larger momentum transfers  and reduce the f.s.i. between the constituents  of the SRC due to production of a state of much larger invariant mass. These are essentially the same processes which one would need to study further color transparency phenomena in hadronic interactions.

\section{Color transparency phenomena current status}
Color transparency (CT)  phenomenon plays  a dual role. It 
 probes both the high energy dynamics of strong interaction  and  the  minimal small size  components of the hadrons. At  intermediate energies it also provides an unique probe of the space time evolution of wave packages which is 
 relevant for interpretation of RHIC  AA data.
 
 The basic tool of CT is  suppression of interaction of small size color singlet configurations. For a dipole of transverse size $d$ perturbative QCD gives 
 \begin{equation}
\sigma(d,x_N)= {\pi^2\over 3} \alpha_s(Q^2_{eff}) d^2\left[x_NG_N(x_N,Q^2_{eff})  +2/3  x_NS_N(x_N, Q^2_{eff})\right], 
\end{equation}
where $Q^2_{eff} \propto 1/d^2, x_N= Q^2_{eff}/W^2$. The second term \cite{Frankfurt:2000jm}
is due to the contribution of quark exchanges which is important for intermediate energies.

There are two critical requirements for the presence of the CT phenomenon - {\it squeezing}  - selection of small size configurations, and {\it freezing}  - selection of high enough energies  to allow squeezed 
configuration to live long enough.

At high energies one can select CT processes by selecting special final states:  diffraction of pion into two high $p_t$  jets, or a small initial state $\gamma_L^*$ - exclusive production of mesons.
QCD factorization theorems were proven  for these processes  based on the CT property of QCD.
CT was observed  dijet production
\cite{Aitala:2000hc} confirming predictions of  \cite{Frankfurt:1993it}.
 The HERA data on vector meson production are also well described by the theory.

Possible ways to look for CT at intermediate energies are reactions 
$A(e,e'p), \gamma_L^* A \to "meson" +A^*$ as well as 
large angle (t/s = const) two body processes:  $a+ b \to c+ d$, see
 Refs. \cite{Jain:1995dd,Miller:2007zzd,Strikman:2007nv} for the recent discussions and detailed references. 
 To observe the CT effect at intermediate energies it is necessary 
 ensure that in the studied kinematics   a small size configuration 
 is "frozen" while it propagates distances comparable to internucleon distances. The expansion / coherence length can be estimated as 
 \cite{CTFL}
  \begin{equation}
 l_{coh}= (.3 \div .4 \, fm ) \cdot p_h[GeV/c].
 \end{equation}
 Hence, $p_h \ge \mbox{4 GeV/c}$ is necessary to observe significant CT effects.  The Jlab $(e,e'\pi^+)$  experiment
\cite{:2007gqa}
  found evidence for CT in this kinematics in agreement with predictions of \cite{Larson:2006ge}.
 Thus there is now a direct evidence for the CT phenomenon - presence of the processes dominated by scattering of mesons in small size $q\bar q$ configurations both at high and intermediate  energies.
  Note also, the A-dependence of the transparency  \cite{:2007gqa} provides a support to a rather small value of the coherence length (Eq.~2) for pions (much smaller than the one assumed in the models of the heavy ion collisions).
   
  Situation in the nucleon sector is more complicated.   
  The data \cite{Leksanov:2001ui} are consistent with the Glauber calculation of \cite{Frankfurt:1994nn} for    $p_N= $ 6 GeV while freezing effect maybe responsible for increase of transparency  for  $p_N \ge $ 8 GeV/c in line with expectations of \cite{Frankfurt:1994nn}. A drop of transparency at  
   $p_N \ge $ 10 GeV/c maybe due to the    competition between contributions of small and large size configurations, see \cite{Jain:1995dd} and references therein.

   It may be related to the observation that baryons  are much more complicated objects than mesons. For example, in the large $N_c$ limit - mesons remains   simple 
  $q\bar q$ systems though nucleons  become solitons which are  classical objects.  
  
  Future studies of CT phenomena are planned at Jlab at 12 GeV including (e,e'p) experiment for $Q^2$ up to 15 GeV$^2$, pion production experiments, and looking for a precursor of the  the CT - chiral transparency - disappearance of the pion cloud \cite{Frankfurt:1996ai}.

  It is very important to have a parallel program of studies with hadron beams. Such program in principle would be possible both at J-PARC and at FAIR.
 In the case of FAIR, the configuration of the PANDA experiment appears to be especially promising. 
  
  \section{Directions for study of color transparency with hadron beams}
  
  {\it i. Study of the two body processes large angle processes with nucleon targets} 
  
  \vspace{3mm}
  
  Understanding of the large angle exclusive processes: $a+ b\to c+d$ remains one of the challenges for  pQCD. Systematic study of a large variety of reactions is available only for incident momentum of 6 and 9.9 GeV/c and below \cite{White:1994tj}. Analysis \cite{White:1994tj} found that cross sections of the processes where quark exchanges are allowed are much larger, and the energy dependence is roughly consistent 
with quark counting rules. Among the biggest puzzles is the ratio of $\theta_{c.m.}=90^o$ cross sections of $\bar p p $ and $pp$ elastic scattering which is below  4\% at 6 GeV/c. At   face value, it indicates extremely strong suppression of the diagrams with
  gluon exchanges in $t$ channel, though more systematic, more precise studies are clearly necessary.
Another puzzle is the oscillation of the differential cross section of the elastic $pp$ scattering at large $t$ around a smooth quark counting inspired parametrization. Are these oscillations present in any of the $p \bar p$ channels?

It appears that  PANDA will have   excellent acceptance for numerous large angle processes  - from the simplest processes $p\bar p \to p\bar p, \pi \pi, K \bar K$ to the processes of production of multi particle  states: baryon - antibaryon  and   meson pairs, etc. In the case of the proton beam the elastic channel is covered reasonably well, though the channels involving  $\Delta$-isobars, nonresonance $\pi N$ production, etc are practically not known.
Another gap in the knowledge is $pn$ scattering which could be studied using the $^2$H pellets. There is a  suggestion that the measurement of the pn/pp ratio may provide an insight on the SU(6) structure of the nucleon wave function at large x \cite{Sargsian:2009}. Overall, comparing all these channels  in $pN$ and $\bar pN$ scattering may lead to a break through in understanding
hard  two body reactions.

\vspace{3mm}

{\it ii. Basic reactions with nuclear targets}

\vspace{3mm}

Use of the nuclear targets allows to test whether basic two body $a+ N  \to c+d$  reactions 
are dominated by small interquark distances.
Presence of the CT in the pion electroproduction and change of the transparency in $pA\to pp (A-1)$ process for $p_N \ge$ 8 GeV \cite{Leksanov:2001ui} indicate that 
 experiments with antiprotons of energies above $6 \div 8$ GeV where a pair of pions is produced in the final state will be able to check by measuring the transparency ratio, $T=\sigma(\bar p A \to \pi^+\pi^- (A-1))/\sigma(\bar p p \to \pi^+\pi^- )$,
  whether the process of annihilation 
happens at small size configurations. Increase of $T$ with energy will signal onset of the CT regime. Presence of the Jlab data on the pion dynamics (both current and from Jlab upgrade which will happen at about the same time as the completion of FAIR) will allow to 
achieve an unambiguous  interpretation of the data.  
 
Another method of detailed study of CT phenomenon is study of rescattering patterns in the exclusive reaction $p(\bar p) +^2H \to p(\bar p) p  +n $ where neutron is slow \cite{Frankfurt:1996uz}.
 This reaction can be separated from background processes without a need to detect the neutron provided a 1\% level momentum resolution for fast particles is achieved and a veto for the pion production is implemented.
Advantage of this reaction is that one can study in detail pattern of the rescattering with characteristic distances between the centers of about 1 fm, hence strongly  suppressing expansion effects.

\vspace{3mm}

{\it iii. Testing chiral dynamics effects}

\vspace{3mm}

It was suggested recently that one can use soft pion theorems in the hard processes to explore the dynamics of the baryon production by virtual photons in the process $\gamma^*N\to N\pi$ with
 $M_{\pi N}-M_{\pi} -M_N\le $ 100 MeV \cite{Pobylitsa:2001cz}. Similar approach is possible in hadronic processes.
One can explore large angle reactions in  non resonance channels  like  
\begin{equation}
\bar p  (p) A \to  N\pi ( \bar p\pi ) + p + (A-1),
 \end{equation}
 where invariant mass of $N\pi$ is close to the threshold. 
 If  the process proceeds through the three quark stage with subsequent emission of  a pion,  transparency will be the same as in the process without pion emission, though if there is no chiral stage absorption will be much larger.
 Another method to probe presence of the  pion cloud close to the interaction point is to look for the charge exchange processes like production of slow $\Delta$ in the process $p+^2H\to pp + \Delta^0$ \cite{Frankfurt:1996ai}.

\vspace{3mm}

 {\it iv. Probing properties of hadron containing charm quark in the $\bar p A$ collisions}

\vspace{3mm}

Charmonium states can  be produced in the nuclear media in the same resonance reactions as the ones  which are studied in 
with the $^2$H target: $\bar p p \to J/\psi,\chi_c,...$. A possibility to use these processes for studies of CT was first suggested in \cite{Brodsky:1988xz}. A subsequent detailed analysis of the effects of CT and Fermi motion was performed in \cite{Farrar:1989vr}. It was found that CT effects due to squeezing of the incoming antiproton are small since the incident energy is small leading to 
a short coherence length for $\bar p$ and produced charmonium.   Effect of the Fermi  motion is large but is under good control. What is unique about these reaction is that at these energies charmonium is formed very close to the interaction point  $\le 1 fm$. Hence they allow   to check the main premise of the CT that small objects interact with nucleons with small cross sections. Ability to select states of varying size: $J/\psi, \chi_c, \psi'$ 
will allow to check how strength of interaction depends on the transverse size of the system up to the sizes comparable to the pion size.  The filtering of $\bar c c $ configurations of smaller transverse size leads  to the   $A$-dependent polarization of the $\chi$ states \cite{Gerland:1998bz}. Study of this phenomenon is also important for understanding of the dynamics of charmonium production in the heavy ion collisions.

There are several other channels sensitive to the dynamics of charm interaction. Near the peak of  resonance production of $J/\psi$ one can look for production of $D, \bar D, \Lambda_c, \psi'$  in the interactions of $J/\psi$ with nucleons of the target nucleus \cite{Gerland:2005ca}. Similar measurements are possible near $\psi'$ resonance. One can use excitation of the  the charmonium resonances above the $\bar DD $ threshold 
and look for softening of the $D$-meson
 spectrum with A which is a measure of the $D-N$ interaction. Such a measurement would be hardly possible in   any other processes.
 
This work is supported by DOE grant under contract DE-FG02-93ER40771.

\end{document}